\documentstyle[12pt,epsf]{article} 
\newcommand{\be}{\begin{equation}} 
\newcommand{\en}{\end{equation}}
\newcommand{\bea}{\begin{eqnarray}}
\newcommand{\ena}{\end{eqnarray}}

\newcommand{\hbo}{\hbox to 1 true cm {\hfill } } 
\newcommand{\tr}{\hbox{tr}}

\def\dslash{\partial\kern-.5em\slash}
\def\kslash{k\kern-.5em\slash}
\def\pslash{p\kern-.5em\slash}

\begin{document} 
\vglue 1truecm
  
\vbox{ UNITU-THEP-18/96
\hfill June 16, 1997
}
  
\vfil
\centerline{\bf Asymptotic freedom in a scalar field theory on the 
   lattice$^1$} 
  
\bigskip
\centerline{ Kurt Langfeld and Hugo Reinhardt } 
\vspace{1 true cm} 
\centerline{ Institut f\"ur Theoretische Physik, Universit\"at 
   T\"ubingen }
\centerline{D--72076 T\"ubingen, Germany.}
\bigskip
  
\vfil
\begin{abstract}
  
An alternative model to the trivial $\phi^4$-theory of the standard model of 
weak interactions is suggested, which embodies the Higgs-mechanism, 
but is free of the conceptual problems of standard $\phi ^4$-theory. 
We propose a $N$-component, O(N)-symmetric scalar field theory, which is 
originally defined on the lattice. The model can be motivated from 
SU(2) gauge theory. Thereby the scalar field arises as a gauge 
invariant degree of freedom. The scalar lattice model is analytically 
solved in the large $N$ limit. The continuum limit is approached via an 
asymptotically free scaling. The renormalized theory evades triviality, 
and furthermore gives rise to a dynamically formed mass of the scalar 
particle.

\end{abstract}

\vfil
\hrule width 5truecm
\vskip .2truecm
\begin{quote} 
$^1$ Supported in part by DFG under contract Re 856/1--3.  

\end{quote}
\eject

In the standard model of weak interactions~\cite{ch84}, the main 
ingredient for the generation of the masses of $W$- and $Z$-bosons as 
well as fermions is a non-vanishing scalar condensate. 
In the usual Higgs mechanism, the non-vanishing scalar condensate 
shows up during a classical treatment of the scalar sector. However, 
it has been shown that a scalar theory with a $\lambda \phi ^4 $ 
interaction of the scalar fields $\phi $ is trivial at quantum 
level~\cite{aiz81}. The quantum theory is only consistent 
with a zero renormalized coupling $\lambda _R$, if the regulator 
(e.g. the momentum cutoff $\Lambda $) is removed. The situation 
can be most easily understood in the large $N$-limit of $O(N)$-symmetric 
$\phi ^4$-theory~\cite{ab76}. Renormalization enforces 
\be 
\frac{6}{\lambda } \; + \; \frac{1}{16 \pi ^{2} } 
\ln \frac{ \Lambda ^{2} }{ \mu ^{2} } \; = \; 
\frac{6}{ \lambda _{R}(\mu ) } \; , 
\label{eq:1} 
\en 
where $\mu $ is an arbitrary renormalization point. Since the stability 
of the bare action requires $\lambda > 0$, one must require 
$\lambda _R(\mu ) \rightarrow 0$, if the cutoff $\Lambda $ goes 
to infinity. 

In order to evade this undesired situation in the Weinberg-Salam model, 
several proposals have been made: first, the scalar field can have a 
non-vanishing expectation value, and the residual interactions of the 
shifted field vanish~\cite{hua87}. Thus the conventional Higgs mechanism 
is possible. Second, the scalar theory is regarded as an effective theory, 
where the momentum cutoff $\Lambda $ is finite~\cite{da83,hel93}. At energy 
scales near this cutoff, new physics comes into the game. The 
parameter range consistent with the phenomenology of weak interactions 
then impose an upper bound on the Higgs mass $m_h$. Recent investigations 
find $m_h \leq 710 \, \pm \, 60 \, $GeV. If the Higgs mass is close to the 
upper bound, the finite cutoff $\Lambda $ 
is in the range $ 2 \ldots 4 \; m_h$~\cite{hel93}. It seems feasible that 
accelerator experiments in the near future will confirm or rule out 
the finite cutoff scenario. Third, an analytical continuation of 
the coupling strength $\lambda $ to negative values~\cite{ga85} 
allows to remove the regulator in (\ref{eq:1}). 
However, stability of the action then requires complex scalar fields. 
The complex nature of the fields induce imaginary parts into the 
effective potential of the scalar condensate~\cite{la96}. The model 
most likely violates fundamental axioms of quantum field theory, e.g. 
the reflection positivity~\cite{os73}. 

The proposals above are not satisfactory from a field theoretical point 
of view. Searching for alternatives to trivial $\phi ^4$-theory, 
Kuti et al.~introduced higher derivative terms into the action 
of the scalar theory~\cite{kut93,kut94}. They found ultraviolet stable 
fixed points and a Higgs mass in the TeV range, which escapes the 
triviality bounds. A different approach was proposed 
in~\cite{mor94,hal95,mor96}. There, the tree level scalar potential was 
generalized to contain arbitrary powers of the scalar field. 
If these powers are multiplied by appropriate inverse 
powers of the cutoff in order to restore the correct energy 
dimension, the models were observed to be 
renormalizable. Moreover, a continuum of 
renormalization group fixed points were found, some of which correspond 
to asymptotically free interactions~\cite{hal95}. 

These observations 
have a tremendous impact on model building and phenomenology, since 
only non-abelian gauge theories are believed so far to be asymptotically 
free and non-trivial. Such a scalar theory therefore would indeed 
serve as a possible candidate for the scalar sector of the Weinberg-Salam 
model. Unfortunately, it turned out~\cite{mor94,mor96} that most 
of the UV-fixed points, found in~\cite{mor94,hal95}, lead to singular 
effective potentials which are not useful in phenomenology. It was 
concluded in~\cite{mor94,mor96} that the investigations in~\cite{hal95} 
do not lead to new physics at the present stage of approximations. 
In the model proposed by Kuti et al.~\cite{kut93,kut94}, complex ghost 
states appear. The discussions on the role of these states in 
phenomenology are still controversial. 

In this letter, we propose a $N$-component, $O(N)$-symmetric scalar 
lattice model, which is a concrete realization of a renormalizable and 
asymptotically free scalar field theory, which, in addition, possesses 
well-behaved effective potentials, and is free of ghost states. 
Our model can be solved in the large 
$N$-limit (without resorting to further approximations as e.g.\ 
those used in~\cite{mor94,hal95,mor96}). 

The lattice scalar model which we wish to study is described by the 
following generating functional 
\bea 
Z[j,\eta ] &=& \int _{-\infty }^{+\infty } [d\phi _x] \; 
\prod _{ \{ x \} } \left[ 1 + \hbox{erf} (\Psi_x a^2) \right] 
\label{eq:2} \\ 
& \times & 
\exp \left\{ - \sum _x a^4 \, \left[ \frac{1}{2} \phi _x (- \partial ^2 ) 
\phi _x  \, - \, \Psi _x^2 \, - \, \eta_x \phi_x \right] \right\} \; , 
\nonumber \\
\Psi _x &=& \sqrt{ \frac{ 3 N }{ 2 \lambda } } \, \left( 
m^2 + \frac{ \lambda }{ 3 N } j_x - \frac{ \lambda }{ 6N } \phi _x ^2 
\right) \; , 
\label{eq:3} 
\ena 
where $a$ is the lattice spacing, $\partial ^2$ is the lattice version 
of the Euclidean d'Alambert operator and {\tt erf} is the error-function. 
The parameter $m$ plays the role of a bare mass, and $j$ is an external 
source, the meaning of which will be explained below.  The functional 
derivative with respect to $\eta _x$ generates scalar field $\phi _x$ 
insertions in Green's functions. Note that 
if we take the bare coupling strength $\lambda $ to zero, we end up 
with a free scalar field theory. 

\begin{figure}[t]
\parbox{6cm}{ 
\hspace{1cm} 
\centerline{ 
\epsfxsize=6cm
\epsffile{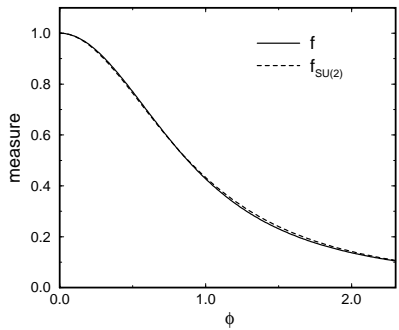} 
}
} \hspace{1cm}
\parbox{7cm}{ 
\caption{ The measure of the scalar functional integral for the 
  case of the SU(2) gauge theory, eq.(8), 
  $ f_{SU(2)} = [1 - \frac{4}{\pi^2} \hbox{ arctan }^2 \phi ]^{1/2}/ 
  (1 + \phi ^2 ) $ and in the coarse grained model, eq.(2), 
  $f = [1 - \hbox{erf} (\phi ^4)] \, \exp \{\phi ^4 \} $. }
} 
\label{fig:1} 
\end{figure} 
Let us motivate the particular type (\ref{eq:2}) of the scalar field 
theory with the ''unusual'' measure. 
Our basic idea is that the scalar degree of freedom naturally 
emerges from Yang-Mills theory as a gauge invariant degree 
of freedom, rather than being introduced by hand as in the electro-weak 
Lagrangian. In fact, such a scalar field theory with the ''non-trivial'' 
integration measure naturally arises both in the lattice~\cite{pol82} and 
the continuum formulation~\cite{ros80,jon91,hr96}. Let us illustrate this 
in the latter case. The gauge invariant partition function 
of Yang-Mills theory can be generically expressed in terms of the 
functional integral 
\be 
Z \; = \; \int {\cal D} \mu (\omega ) \; \int {\cal D} A_i \; 
\exp \{ -S_{YM} \} \; , 
\label{eq:h1} 
\en 
where $S_{YM}$ is the standard Yang-Mills action (see e.g.~\cite{ch84}) 
and $A_{i=1 \ldots 3}$ denote the spatial components of the 
gauge field. $\omega $ is a time-independent, but space dependent 
element of the Cartan subgroup of the gauge group, and $\mu (\omega ) $ 
denotes the corresponding Haar measure. The integration over the compact 
variable $\omega $ stems from the projection onto physical (i.e. gauge 
invariant) states~\cite{jon91,hr96}. For a SU(2) gauge group, this 
integration can be parameterized by 
\be 
\int {\cal D} \mu (\omega ) = \int _0^\pi {\cal D} \chi \; \sin ^2 \chi 
= \int _{-1}^{+1} da_0 \; \sqrt{ 1 - a_0^2 } \; , \hbo 
a_0 = \cos \chi \; . 
\label{eq:h2} 
\en 
The temporal component $A_0$ of the gauge field, which enters the 
action $S_{YM}$ in (\ref{eq:h1}), is time independent and 
related to $\omega $ by 
\be 
A_0 \; = \; - \frac{1}{T} \, \ln \omega \; , \hbox to 4cm { 
\hfil T: time period } \; , 
\label{eq:h3} 
\en 
which suggests to interpret $\omega $ as the (diagonal) Polyakov line. 
Let us emphasize that only the eigenvalues of the Polyakov line, 
i.e. $\exp \{ \pm i \chi \} $, contribute to the integration (\ref{eq:h2}) 
and that these eigenvalues (and therefore $\chi $) are gauge 
invariant. 

The representation (\ref{eq:h1}) also straightforwardly follows from 
continuum limit of the lattice formulation as well as 
from the standard continuum formulation~\cite{hr96} with 
Faddeev-Popov gauge fixing in the gauge 
\be 
A_0^{ch} =0 \; , \hbo \partial _0 A_0^n =0 \; , 
\label{eq:h4} 
\en 
where $A_0^n$ and $A_0^{ch}$ denote the ''neutral'' (diagonal) 
and the ''charged'' (off-diagonal) part of the gauge field. 
In this case the Haar measure arises from the Faddeev--Popov determinant. 

Let us show that the theory (\ref{eq:h1}) of the gauge invariant 
variable $a_0$ has a similar structure as the model (\ref{eq:2}) of the 
scalar field. Assume that the spatial components $A_i$ in (\ref{eq:h1}) 
are integrated out. This procedure would yield an effective theory of 
the scalar $a_0$, which is equivalent to the original Yang-Mills theory and 
which is hence asymptotically free and non-trivial. This simple argument 
shows that asymptotically free and non-trivial effective scalar theories 
indeed exist.

In order to introduce the standard support of a scalar field integral, 
we define $a_0(x) = 2/\pi \, \hbox{arctan} \, \phi (x) $, which maps 
the variable $a_0 \in [-1,+1]$ onto $\phi (x) \in ]-\infty , \infty [$. 
The $a_0$-integral in (\ref{eq:h2}) then results in 
\be 
\int _{-\infty }^{\infty } {\cal D} \phi \; 
\frac{2}{\pi } \frac{ \sqrt{ 1 - \frac{4}{\pi^2} \hbox{ arctan }^2 
\phi } }{ 1 + \phi ^2 } \; \ldots \; . 
\label{eq:e3} 
\en 
Figure 1 compares the measure of the $\phi $ functional integral 
(\ref{eq:e3}) with that of our model (\ref{eq:2}), where we have set 
$j$ and $m$ to zero. For practical applications (e.g. a Monte-Carlo 
simulation), both curves are the same. The considerations above 
illustrate that the measure of the scalar functional integral (\ref{eq:2}) 
can be understood as a relict of the integration over the compact 
variable $\omega $, which ensures the proper projection onto physical 
states contributing to the Yang-Mills partition function. 
Thus the scalar field of our model can be interpreted as 
a gauge invariant degree of freedom of Yang-Mills theory. 
In this letter, we do not further 
pursue the rigorous construction of the scalar model from gauge 
theory, but rather study the coarse grained model (\ref{eq:2}) 
which is exactly solvable in the large $N$-limit. 

In the naive continuum limit $(a \rightarrow 0)$, the generating functional 
(\ref{eq:2}) takes the form 
\be 
Z_{naive} [j=0, \eta =0 ] \; = \;  \int _{-\infty }^{+\infty } [d\phi _x] \; 
\exp \left\{ - S_{naive} 
\right\} \; , 
\label{eq:4} 
\en 
where (for illustrative purposes, we here choose $m=0$ and $j=0$) 
\be 
S_{naive} = \int d^4x \, \left\{ \frac{1}{2} \phi _x (- \partial ^2 ) 
\phi _x \, + \, \frac{\widetilde{m}^2}{2} \phi^2_x \, - \, 
\frac{\widetilde{\lambda }}{N} \phi^4_x 
\right\} \; + \; {\cal O }(a^2) \; , 
\label{eq:5} 
\en 
where 
\be 
\widetilde{m}^2 \; = \; \sqrt{ \frac{2 \lambda }{3 N \pi } } \, 
\frac{1}{a^2}  \; , 
\hbo 
\widetilde{\lambda } \; = \; \frac{\lambda }{ 24 } 
\left(1-\frac{2}{\pi} \right) \, 
\hbo 
\label{eq:6} 
\en 
Thus the naive continuum limit corresponds to a continuum $\phi ^4$-theory 
with negative quartic interaction. Nevertheless, the full lattice model 
(\ref{eq:2}) is well-defined, since the integrand in (\ref{eq:2}) 
behaves like $1/\phi^2_x a^2$ for $\phi^2_x a^2 \gg 1$. This implies that 
precisely the terms which vanish in the naive continuum limit 
stabilize the lattice action. 
Below, we will show that the above lattice model possesses 
at quantum level an interesting continuum limit, which is non-trivial 
in the sense that the scalar fields have non-vanishing vacuum 
expectation values. Furthermore, the continuum limit is approached via an 
asymptotically free scaling, i.e. $\lambda \rightarrow 0$ for $a 
\rightarrow 0 $. One can already anticipate this remarkable scaling 
behavior, by starting from (\ref{eq:4}) and using standard perturbation 
theory. In leading order, our model essentially represents standard 
$\phi ^4$-theory with, however, negative quartic coupling (see (\ref{eq:6})). 
Due to this inverse sign of the coupling, the renormalization group 
$\beta $-function of our model is negative implying asymptotic freedom. 
In order to get access to the non-trivial vacuum properties, we do 
not further pursue perturbation theory, but will study the model 
in the large $N$-limit.

Let us first discuss the classical field, which is produced by 
the functional derivative of $Z[j]$ with respect to the external 
source $j$, i.e. 
\be 
{\cal C } \; := \; - \, \frac{\delta }{ \delta j(x) } \, \ln Z[j] \, 
\vert _{j=0}. 
\label{eq:6a} 
\en 
In the naive continuum limit, this classical field basically measures 
$\langle \phi ^2 \rangle $ (up to a constant shift) 
\be 
{\cal C} \; = \; \frac{\lambda }{6N} \left( 1- \frac{2}{\pi} \right) 
\, \langle \phi ^2 \rangle \; - \; \sqrt{ \frac{ 2 \lambda }{ 3 \pi N }} 
\frac{1}{a^2} - \left( 1- \frac{2}{\pi} \right) m^2 \; + \; 
{\cal O} (a^2) \; . 
\label{eq:6b} 
\en 
For ${\cal C} =0$, the scalar condensate is proportional 
to the mass parameter, as it is the case in a classical field 
theory with spontaneous symmetry breaking at tree level. 
Therefore, any non-vanishing value of ${\cal C}$ 
measures the non-trivial scalar condensate which occurs in the 
ground state of the quantum theory (for vanishing renormalized mass). 

In the following, we will solve the lattice model (\ref{eq:2}) in the large 
$N$ limit. For this purpose, we introduce the auxiliary field $M_x$ by 
\be 
\prod _{\{x\}} \left[ 1 \, + \, \hbox{erf} \left( \Psi _x a^2 \right) 
\right] \; = \; 
\int _0^\infty [dM_x] \; 
\exp \left\{ - \sum _{\{x\}} a^4 \, \left( 
\sqrt{ \frac{3N}{2 \lambda }}  M_x -\Psi _x \right)^2 \right\} \; , 
\label{eq:7} 
\en 
where we have absorbed an unimportant factor into the measure of $M_x$. 
Inserting (\ref{eq:7}) into the definition of the model (\ref{eq:2}), 
a Gaussian integral over the scalar fields is left. Performing this 
integration, one finds 
\bea 
Z[j,\eta ] &=& \int _0^\infty [dM_x] \; \exp \left\{ -S_M \right\} \; , 
\label{eq:8} \\ 
S_M &=& \frac{1}{2} \tr \ln ( - \partial ^2 + M_x ) \; + \; 
\sum_x a^4 \left( \frac{3N}{2 \lambda } M_x^2 \, - \, \frac{3N}{ \lambda } 
m^2 M \, - \, j_x M_x \right) 
\label{eq:9} \\ 
&-& \frac{1}{2} \sum _{xy} a^4 \, \eta _x ( - \partial ^2 + M_x )^{-1}_{xy} 
\eta _y \; . 
\nonumber 
\ena 
This model is completely equivalent to that in (\ref{eq:1}). The latter 
version (\ref{eq:8}), however, is more suitable to perform the 
large $N$-expansion, since fluctuations of the field $M_x$ are suppressed 
by powers of $1/N$.  In order to study the continuum limit, we carefully 
investigate the divergences of the trace term in (\ref{eq:9}) which 
will occur, if we shrink the lattice spacing to zero. 
Assuming a constant field $M$, this trace term is given by 
\be 
N \int _{-\frac{\pi}{a} }^{+\frac{\pi}{a} } \frac{d^4k}{(2\pi )^4} \, 
\ln \left( 2 \sum _{\mu =1 }^4 (1 - \cos k_\mu a )  +  Ma^2 
\right) \, = \, - \frac{N}{a^4} \int _0^\infty \frac{ds}{s} 
e^{-s Ma^2 }  e^{-8s} \left[ I_0(2s) \right]^4 \; , 
\label{eq:10} 
\en 
where we have factored out the lattice volume and have dropped a 
$M$-independent constant. $I_0(x)$ is the Bessel function of the first 
kind. Inserting (\ref{eq:10}) in (\ref{eq:9}), one observes that the 
quadratic divergence, i.e.~$\propto 1/a^2$, can be absorbed in the 
bare mass $m$. The logarithmic divergence of (\ref{eq:10}) turns out 
to be proportional to $M^2$ implying that (\ref{eq:9}) is renormalized 
by setting 
\be 
\frac{3}{2 \lambda (\Lambda ^2) } \, - \, \frac{1}{64 \pi ^2} 
\ln \frac{ \Lambda ^2 }{\mu ^2 } \; = \; \frac{3}{2 \lambda _R (\mu ) } 
\; , \hbo 
\Lambda ^2 := 1/a^2 \; , 
\label{eq:11} 
\en 
where $\mu $ is an arbitrary reference scale, and $\Lambda $ is the 
momentum cutoff provided by the lattice spacing. Equation (\ref{eq:11}) 
is of crucial importance. Firstly, it tells us, that the cutoff 
$\Lambda $ can be taken to infinity without enforcing a vanishing 
renormalized coupling strength $\lambda _R$. Hence, the model evades 
triviality, which is inherent in standard $\phi ^4$-theory. Secondly, 
the theory (\ref{eq:2}) is asymptotically free, which can be easily 
verified by calculating the renormalization group $\beta $-function, 
which in the large $N$ limit is 
\be 
\beta (\lambda ) := \Lambda  \frac{ d \lambda (\Lambda ) }{ d \Lambda } 
\; = \; - \, \frac{ \lambda ^2 }{48 \pi ^2 } \; . 
\label{eq:12} 
\en 
One might worry that the renormalized coupling diverges at a certain 
momentum at low energies. This behavior is generic for a large 
class of asymptotically free theories such as QCD and the 
Gross-Neveu model. In the latter model, it was observed that the 
divergence at low momenta is an artifact due to the assumption of a 
perturbative vacuum, and that the divergence is in fact screened by 
a mass which is dynamically formed in the true vacuum~\cite{la95}.

In order to demonstrate that the model (\ref{eq:2}) possesses non-trivial 
ground state properties, we present the effective potential $U$ for the 
composite field ${\cal C}$. This potential can be obtained by a 
Legendre transformation of $Z[j,\eta =0]$ with respect to $j$ thereby 
assuming constant classical fields ${\cal C}$. In the large $N$ limit, 
the final result is for the case of a vanishing renormalized mass 
(details will be presented elsewhere) 
\be 
U({\cal C}) \; = \; \frac{N}{ 64 \pi ^2 } \, {\cal C}^2 \, \left( 
\ln \frac{ {\cal C} }{ M_0 } \, - \, \frac{1}{2} \right) \; , 
\label{eq:13} 
\en 
where $M_0$ is a renormalization group invariant scale, which arises from 
dimensional transmuting the coupling strength $\lambda _R$. 
Functional derivatives of the effective action with respect to 
the classical field provide vertex functions which contain insertions 
of the composite field ${\cal C}$. Differentiations of the effective 
potential with respect to constant fields yield vertex functions at 
zero momentum transfer. From (\ref{eq:13}), we learn that our model 
possesses a tower of non-vanishing vertex functions, the energy scale 
of which is set by the renormalization group invariant scale $M_0$. 
Our model therefore exhibits non-vanishing interactions in the continuum 
limit (non-triviality).  

Stationary points ${\cal C}_0$ of the potential (\ref{eq:13}) 
immediately get a physical interpretation, if we calculate the propagator 
of the scalar particle $\phi _x$. 
Taking twice the functional derivative of $Z[j=0,\eta ]$ with respect 
to $\eta $ (and setting $\eta =0 $ afterwards\footnote{This ensures 
that $U({\cal C})$ is stationary at the corresponding ${\cal C}$ value.}), 
this propagator is in leading order of the large $N$-expansion 
\be 
S_{xy} \; = \; \frac{1}{ - \partial ^2 \, + \, {\cal C}_0 } \; . 
\label{eq:14} 
\en 
We therefore identify ${\cal C}_0$ as the mass squared of the scalar 
particle. 

On the other hand, stationary points of the effective potential correspond 
to possible candidates for the vacuum. Each stationary point describes 
a phase of the model. The phase with minimal potential $U$ 
constitutes the ground state at zero temperature. In the case of the 
potential (\ref{eq:13}), two phases are present. In the phase characterized 
by ${\cal C}_0=0$, the scalar particle is massless. However, the second 
phase (provided by the stationary point at ${\cal C}_0= M_0$)
describes the true ground state, since this state has a lower 
vacuum energy density, i.e. $- M_0^2/128\pi^2 $ (note that the 
perturbative state has by definition zero energy density). 
We call this phase 
non-trivial, since it exhibits a dynamically formed mass of the 
scalar particle. In the latter phase, the ratio of the dynamical mass 
squared and the vacuum energy density is in leading order of the 
large $N$-expansion 
\be 
\sqrt{N} \frac{ M_0 }{ \sqrt {-U({\cal C}_0=M_0) } } \; = \; 8 \sqrt{2} 
\, \pi \; . 
\label{eq:15} 
\en 
To estimate the order of magnitude of the scalar mass, we set 
$N=1$ and choose a generic value of the vacuum energy density of 
the theory of weak interactions, i.e. $U({\cal C}_0=M_0) \approx -
(200 \, \hbox{GeV})^4$. From (\ref{eq:15}), we then find a mass of the 
scalar particle quite above $1 \, $TeV. This result supplements those 
obtained in~\cite{kut94}, where a Higgs mass in the 
TeV range was also reported in a different type of scalar field theory. 

It was first shown by Kuti et al.~\cite{kut93,kut94} that certain scalar 
field theories possess ultraviolet fixed points, and that the Higgs mass 
predicted by these theories evade the triviality bounds. 
It was argued by renormalization group techniques 
first by Morris~\cite{mor94} and independently by Halpern and 
Huang~\cite{hal95} that certain non-polynomial scalar field theories 
which contain the momentum cutoff at tree level are non-trivial 
and asymptotically free. In this letter, we have provided for the first time 
an explicit example for such a theory. We have proposed an O(N)-symmetric 
scalar field theory on the lattice which is solvable in the large $N$-limit. 
The key point is that the functional integration over the scalar field 
is supplemented with a non-trivial measure. We have argued that such 
a measure naturally arises, if the scalar field is interpreted 
as effective gauge invariant degree of freedom of SU(2) 
gauge theory. The quantum model of our toy theory 
has a continuum limit and is asymptotically free. 
The model exhibits two phases. In the vacuum phase, the scalar particle 
possesses a dynamically generated mass in the TeV range. The results therefore 
signal non-triviality.

\bigskip 
{\bf Acknowledgments: } 

Many discussions with L.~v.~Smekal on the issue of triviality of 
standard $\phi ^4$-theory are greatly acknowledged.

\end{document}